\documentclass{elsart}
\usepackage{graphicx}
\usepackage{amssymb}

\begin{document}

\begin{frontmatter}

\title{Multifractal Behavior of the Korean Stock-market Index KOSPI}

\author[jwl,par1]{Jae Woo Lee\corauthref{cor1}}
\ead{jaewlee@inha.ac.kr}
\author[jwl]{Kyoung Eun Lee}
\author[par1,par2]{Per Arne Rikvold}
\corauth[cor1]{Corresponding author. Tel.: 82+32+8607660; fax: 82+32+8727562}
\address[jwl]{Department of Physics, Inha University, Incheon 402-751, Korea}
\address[par1]{School of Computational Science,  
Florida State University, Tallahassee FL 32306-4120, USA}
\address[par2]{Department of Physics, Center for Materials Research \\
and Technology, and National High Magnetic Field Laboratory, 
Florida State University, Tallahassee, FL 32306-4350, USA}

\begin{abstract}

We investigate multifractality in the Korean stock-market index KOSPI. 
The generalized
$q$th order height-height correlation function shows  multiscaling
properties. There are two scaling regimes with a crossover time around $t_c 
=40$~min. We consider the original data sets and the modified data
sets obtained by removing the daily jumps, which occur due to the difference
between the closing index and the opening index. To clarify the origin of
the multifractality, we also smooth the data through convolution
with a Gaussian function. After convolution we observe that the
multifractality disappears in the short-time scaling
regime $t<t_c$, but remains in the long-time scaling regime
$t>t_c$, regardless of whether or not the daily jumps are removed. We suggest
that multifractality in the short-time scaling regime is caused 
by the local fluctuations of the stock index. But the multifractality
in the long-time scaling regime appears to be due to the intrinsic trading properties,
such as herding behavior, information outside the market, the
long memory of the volatility, and the nonlinear dynamics
of the stock market.
\end{abstract}

\end{frontmatter}

\newcommand{\be}{\begin{equation}}
\newcommand{\ee}{\end{equation}}

\section{Introduction}
In recent years,
concepts and techniques from statistical physics have been widely applied 
to 
economics~\cite{MA99,MA97,BP00,SO03,MS95,BS94,BA1900,MA63,FA63,GP99,GM99,LK99,GP00,SA00,KK02,BC03,NA02,LE02,LL04},
and the complex behaviors of economic systems
have been found to be very similar to those of
complex systems customarily studied in statistical physics. 
Stock-market indexes around
the world have been accurately recorded for 
many years and therefore represent a rich source of data for quantitative
analysis, and  the statistical behaviors
of stock markets have been studied by various methods, such as
distribution functions~\cite{GP99,GM99,LK99,LL04},
correlation functions~\cite{LK99,GP00,SA00}, 
multifractal analysis~\cite{LL04,GD02,WH01}, and
network analysis of the market structure~\cite{KK02,BC03}.

Multiscaling properties have been reported for many economic time 
series~\cite{KC04,SK00,KY03,AR02,XX03,EK04,BE03,AI02,IA99,TP03}.
Multifractality has been observed in stock 
markets~\cite{GD02,AI02,KA00,BE01,MA03}, the price of crude 
oil~\cite{AR02}, the price of commodities~\cite{MA03},
and foreign exchange rates~\cite{KY03,XG03}.
In daily stock indexes and foreign exchange rates,
the generalized Hurst exponent  $H_q$ decreases monotonically with 
$q$~\cite{KY03,AI02,KA00,BE01,MA03,XG03} (see formal definitions
of $H_q$ and $q$ in Eqs. (1) and (2)
below).
In the U.S. NASDAQ index, two scaling regimes have been
reported: one quasi-Brownian and the other 
multifractal~\cite{BE01}. 
Two scaling domains have also been reported  in the
fluctuations of the price of crude oil: $H_q$ increases
with $q$ in the short-time domain, but decreases with
$q$ in the long-time domain~\cite{AR02}.
The existence of multiple scaling regimes seems to depend on the resolution
of the data sets in the economic time series.

Although many economic time series display
multifractality (in the theory of surface scaling analysis
referred to as {\em multiaffinity}),
the origins of multifractality in the
stock market are not well understood. 
It has been suggested that herding behavior and nonlinear
complex dynamics of the stock market induce multiscaling~\cite{GP99}.
However, it is very difficult to quantify herding behavior and complex 
dynamics in the stock market. 
Buend{\'\i}a {\em et al.} observed  multiaffinity
in a frustrated spring-network model simulating the surface structure
of cross-linked polymer gels~\cite{BM02}. Removing vertical
discontinuities from the rough surface by convolution with a Gaussian, 
they observed that the multiaffine surface changed into a self-affine one~\cite{BM03}.
They concluded that vertical discontinuities can be one cause of
multiaffinity.
Mitchell conversely introduced artificial vertical discontinuities into a
self-affine surface and observed that the surface became
multiaffine~\cite{MI02}. 

In the present paper we investigate
multifractality in the Korean stock-market
index KOSPI (Korean Composite Stock Price 
Index). We observe that local fluctuations, including jumps, are 
responsible for multifractality in the short-time scaling regime. 
However,  multifractality
in the long-time scaling regime is not removed by smoothing
of the time series.

\section{Method and Results}
We consider a set of data recorded every
minute of trading from March 30, 1992, through November 30, 1999. 
We count the time during trading hours and remove closing hours,
weekends, and holidays from the data. 
Denoting the stock-market index as $x(t)$, the generalized $q$th
order height-height correlation function~(GHCF) $F_q (t)$ is defined by

\be
F_q(t) = \langle \vert x(t^{\prime} +t) - x(t^{\prime}) \vert^{q} \rangle^{1/q},
\ee
where the angular brackets denote a time average over the time series.
The GHCF $F_q (t)$ characterizes the correlation properties 
of the time series $x(t)$, and for a multiaffine series a power-law behavior like
\be 
F_q(t) \sim t^{H_q}
\ee
is expected,
where $H_q$ is the generalized $q$th order Hurst exponent~\cite{MK}. If $H_q$
is independent of $q$, the time series is monofractal. If $H_q$ depends
on $q$, the time series is multifractal.
Multifractality is a distinctive property of the stock market index
observed. 

\begin{figure}
\includegraphics[width=13.5cm,height=17cm,angle=0,clip]{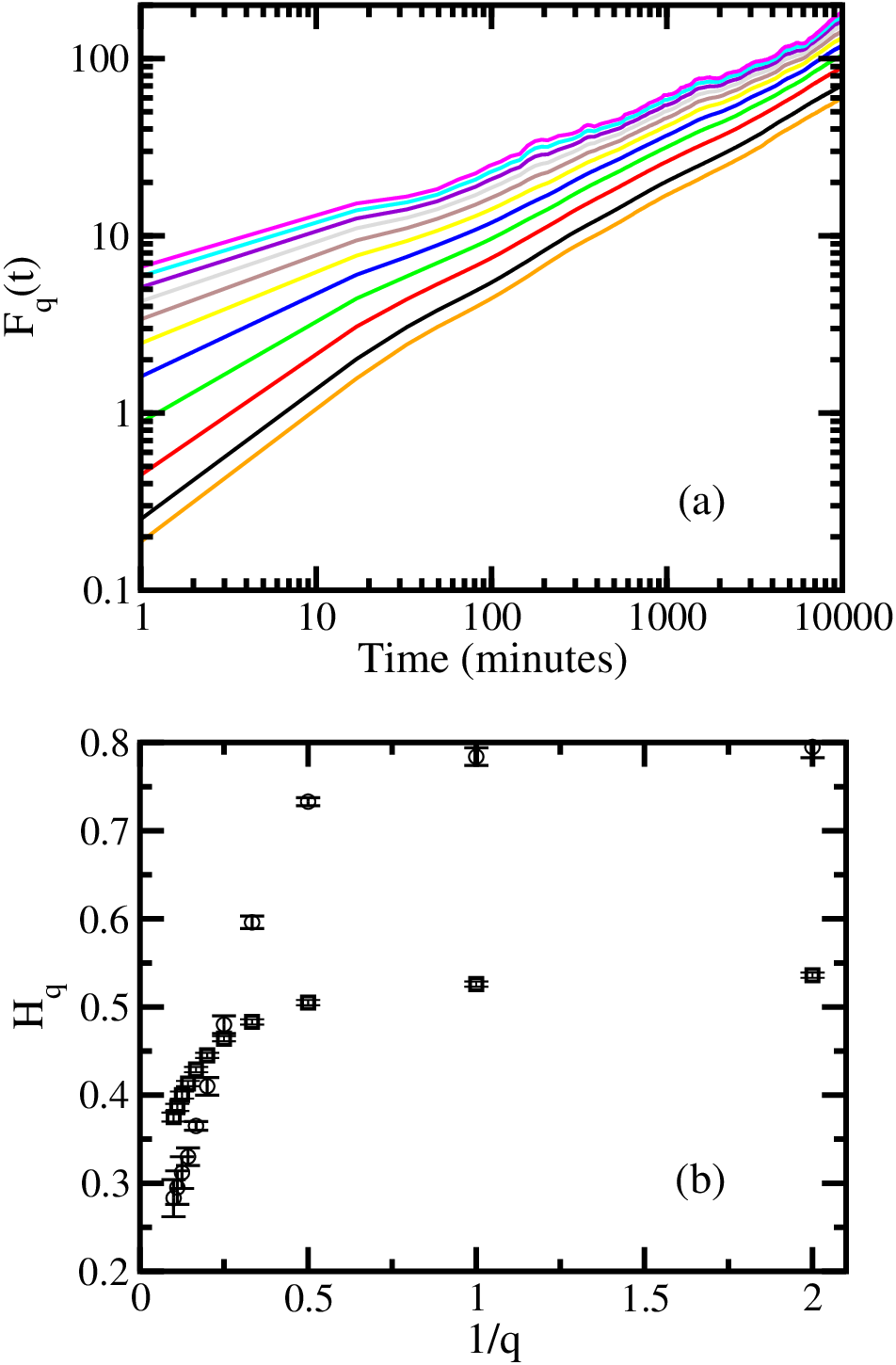}
\caption[0]{
(Color online.)
(a) Log-log plot of the generalized height-height correlation function $F_q (t)$ versus
time $t$ for the prices of the Korean stock-market
index, KOSPI with $q=0.5, 1, 2, \cdots, 10$
from bottom to top. 
(b) The generalized Hurst exponents $H_q$ versus $1/q$ for the
price of KOSPI in the scaling regimes $t<t_c ~(\circ)$ and $ t > t_c ~(\square)$
where $t_c \simeq 40$~min. The error bars here and in Figs. ~3(b) and 4(c) are
standard deviations in the slopes of least-square fits to 
log-log plots of $F_q (t)$.
}
\label{Fig1}
\end{figure}

In Fig.~\ref{Fig1}(a)  we present the GHCF as a function of the time
interval $t$. We observe clear multifractal behavior in the 
time series of  the stock index.
There are two different scaling regimes separated by a crossover time 
of about $t_c =40$~min. 
The slopes of the log-log plot depend on $q$ 
in each scaling regime.
As shown in Fig.~1(b), $H_q \sim 1/q$ for large $q$ in both scaling regimes,
while it saturates for small $q$. The $1/q$ behavior is consistent with 
numerical observations in Refs.~\cite{BM03,MI02} and analytical results
for functions that include discontinuities in Ref.~\cite{MI02}.
These behaviors are
different from the multifractality in the price of crude oil~\cite{AR02}. 
Alvarez-Ramirez {\em et al.}~ reported the existence of two 
scaling regimes for crude-oil prices~\cite{AR02}, 
but $H_q$ {\em increased} with $q$ in the long-time scaling regime.
Kim {\em et al.} 
did not observe a short-time scaling regime for the daily 
Yen-Dollar exchange rate~\cite{KC04}.

\subsection{Removal of daily jumps}
 We analyzed the KOSPI time series to find the origin of  multifractality
in the Korean stock market. Inspired by Refs. \cite{BM02,BM03,MI02},
we removed the daily jumps of the stock index 
due to the difference between the closing and opening index. 
In KOSPI, there was no trading after closing and before 
opening the market until 1995. So, the changes of the index 
associated with
the daily jumps are small during those years. Since 1995, after-market trading
and before-market trading have been allowed for one hour each, leading to
substantial overnight jumps. 
There are additional big daily changes
during the period of the Asian financial crisis around November, 1997. 
In Fig.~2, we present three time series: the original time series, the
time series with the daily jumps removed, and the
time series of the daily jumps only.
In the period between 1992 and 1999, there were bubbles in
the Korean stock market around 1995, and there were big crashes
around 1997 after the Asian financial crisis. 
Larger daily changes can be observed around the period of
the Asian financial crisis. 

\begin{figure}
\includegraphics[width=12.5cm,height=11cm,angle=0,clip]{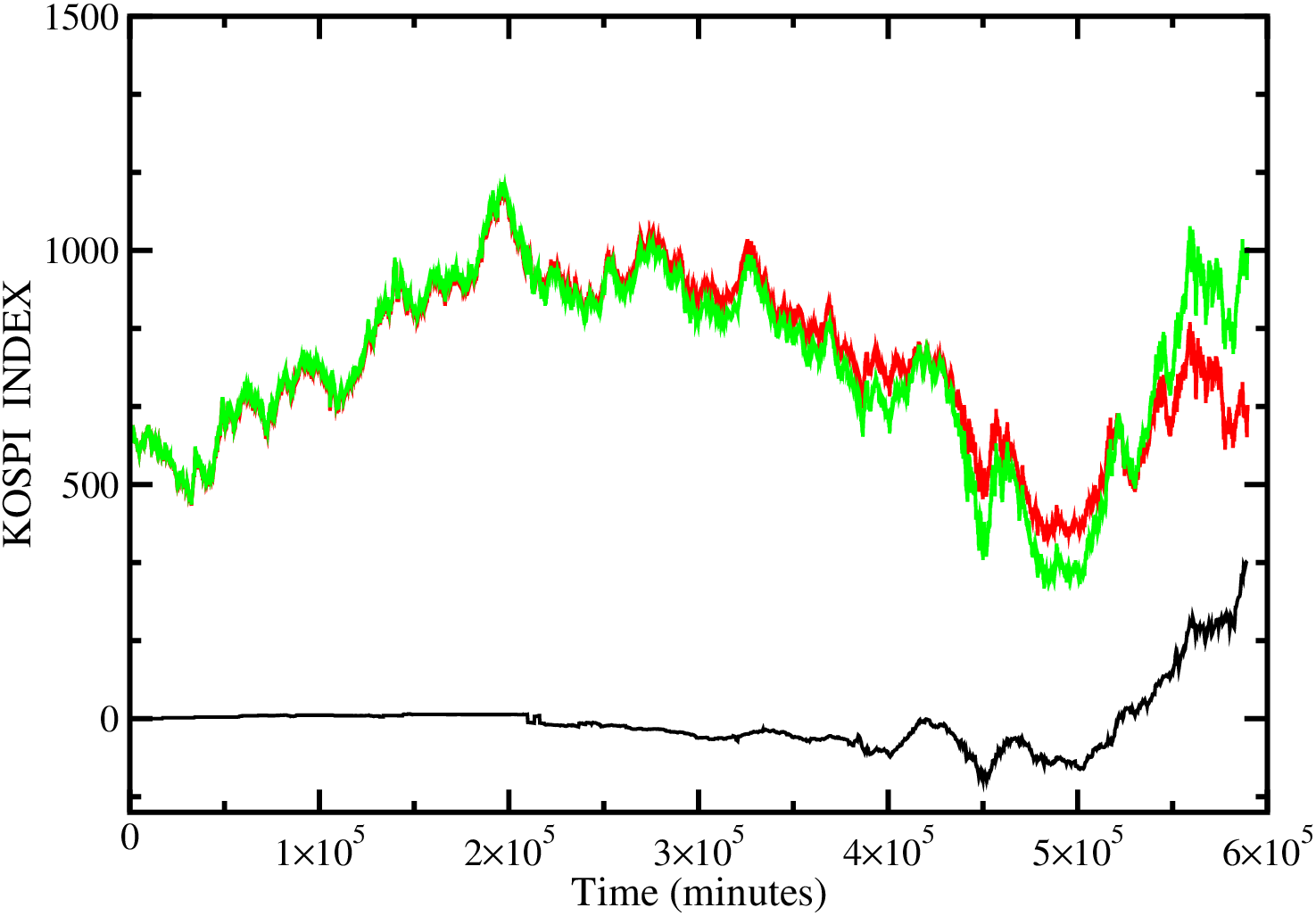}
\caption[0]{
(Color online.) The time series of the Korean stock-market index, KOSPI: 
original one-minute time series~(light gray, green online), the time
series with the daily jumps removed~(dark gray, red online), and the time series
of the daily jumps only~(black).
}
\label{fig2}
\end{figure}

\begin{figure}
\includegraphics[width=13.5cm,height=17cm,angle=0,clip]{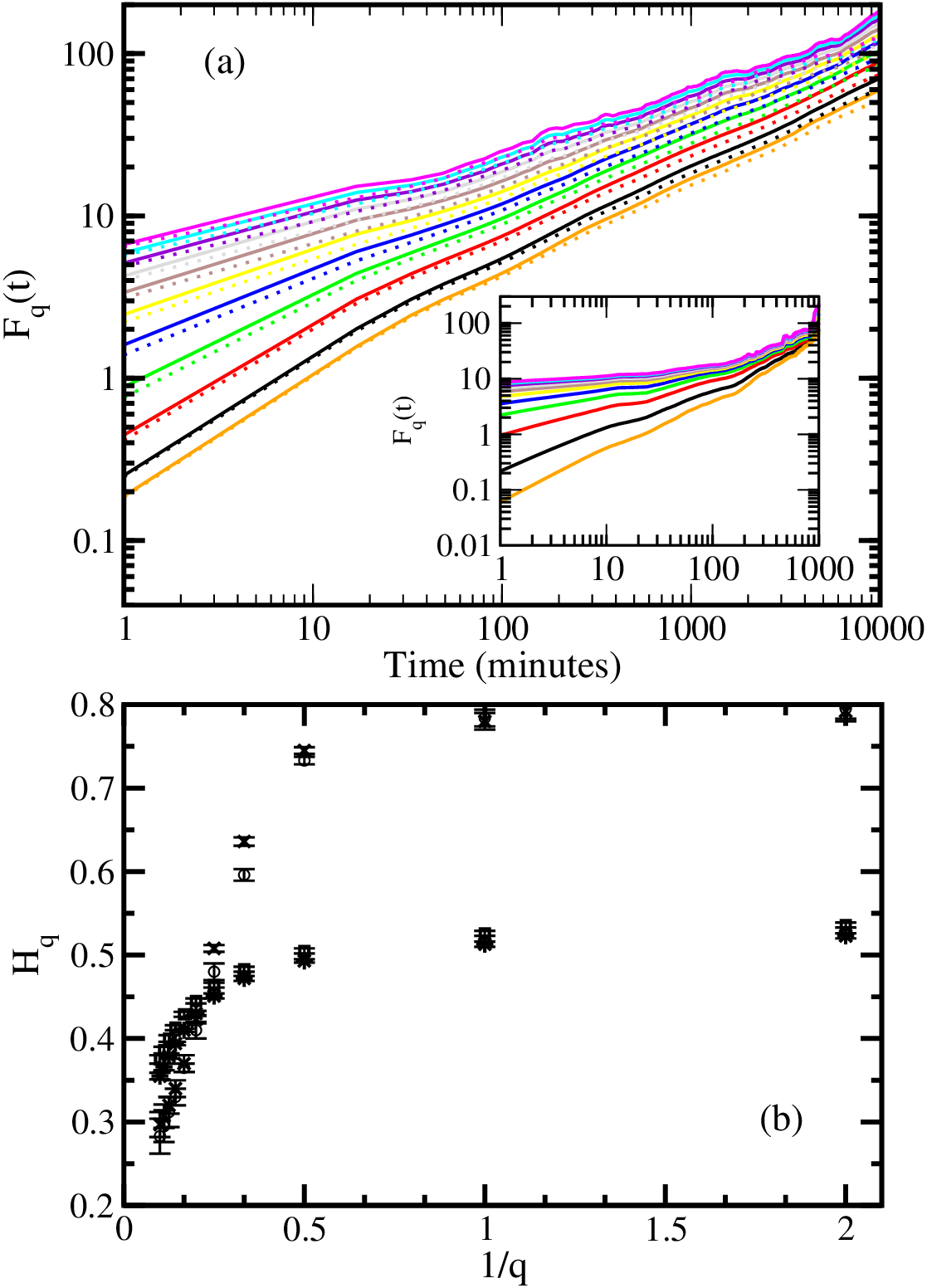}
\caption[0]{
(Color online.) (a)  Log-log plots of the generalized 
height-height correlation function $F_q (t)$ 
versus time in the time series of 
the original index~(solid lines) and the
index with the daily jumps removed~(dotted lines) for $q=0.5, 1, 2, \cdots, 10$
from bottom to top. Inset: the generalized height-height correlation
function versus time for the daily jumps.
(b) The generalized Hurst exponents $H_q$ versus $1/q$ 
for the index with the daily jumps removed
in the scaling regimes $t<t_c ~(\times)$ and $ t > t_c ~(*)$
where $t_c \simeq 40$~min. 
For easy comparison we also include
the original index in the scaling regimes $t<t_c ~(\circ)$ and $ t > t_c ~(\square)$
(identical to Fig.~1(b)).
}
\label{fig3}
\end{figure}

  We show the GHCF for the original KOSPI, the stock index with the 
daily jumps removed, and the time series of the daily jumps only in Fig.~3(a).
In the short-time scaling regime $t<t_c$, the deviations between the original
and the modified data sets increase with increasing $q$. But in the
long-time scaling regime $t>t_c$, the deviations are larger 
for small $q$. For large $q$, the GHCF have large fluctuations,
and there may be additional scaling regimes for very long times.
However, practically no change is observed in $H_q$, as shown in  
Fig.~3(b).
Removing the daily jumps thus does not delete the multifractal properties of 
the stock index and only slightly changes the generalized Hurst exponents.
In the inset in Fig.~3(a)
we present the GHCF of the time series of the daily jumps.
We observe obvious multifractality, and there are several scaling regimes.

\subsection{Gaussian smoothing}
Next, as simply removing the discontinuities has little effect,
 we consider the effects 
of smoothing the stock-index time series
by convolution with a Gaussian of standard deviation 5 minutes. 
The convolution of two functions $f(t)$ and $g(t)$ is defined
by
\begin{equation}
\tilde{f}=\int_{-\infty}^{\infty} f(t) g(t-u) du,
\end{equation}
where $g(t)$ is a Gaussian convolution kernel~\cite{NR}. 
In discrete time series the convolution is a sum instead of an integral 
\begin{equation}
\tilde{f}_i = \sum_{j=1}^{m} f_j g_{i-j}.
\end{equation}
In Fig.~4(a) we show
a short segment of the original time series and 
the time series smoothed by convolution with a Gaussian.
Although this procedure removes short-time fluctuations of the stock index,
global trends remain.
In Fig.~4(b) we show  the GHCF 
before and after the convolution.  Before the convolution, the GHCF
shows multifractality.
After the convolution we observe monofractal
behavior in the short-time scaling regime, $t<t_c$.
In this scaling regime, the lines of the log-log plots are all parallel, 
with the same slope, $H_q (t < t_c ) =0.986(4) \simeq 1$, independent
of  $q$, as shown in table~\ref{hst}. 
$H_q =1$ means that the convoluted time series is a smooth curve. 
The result of $H_q \simeq 1$ for $t< t_c$ and unchanged $H_q$ for 
$t>t_c$ agrees with those obtained from Gaussian smoothing of  simulated
hydrogel surfaces in Ref.~\cite{BM03}.
For $t>t_c$, the convoluted time series yield the same slopes as the 
original time series. The corresponding values of $H_q$ are shown vs $1/q$
in Fig.~4(c).

\begin{figure}
\includegraphics[width=13.5cm,height=17cm,angle=0,clip]{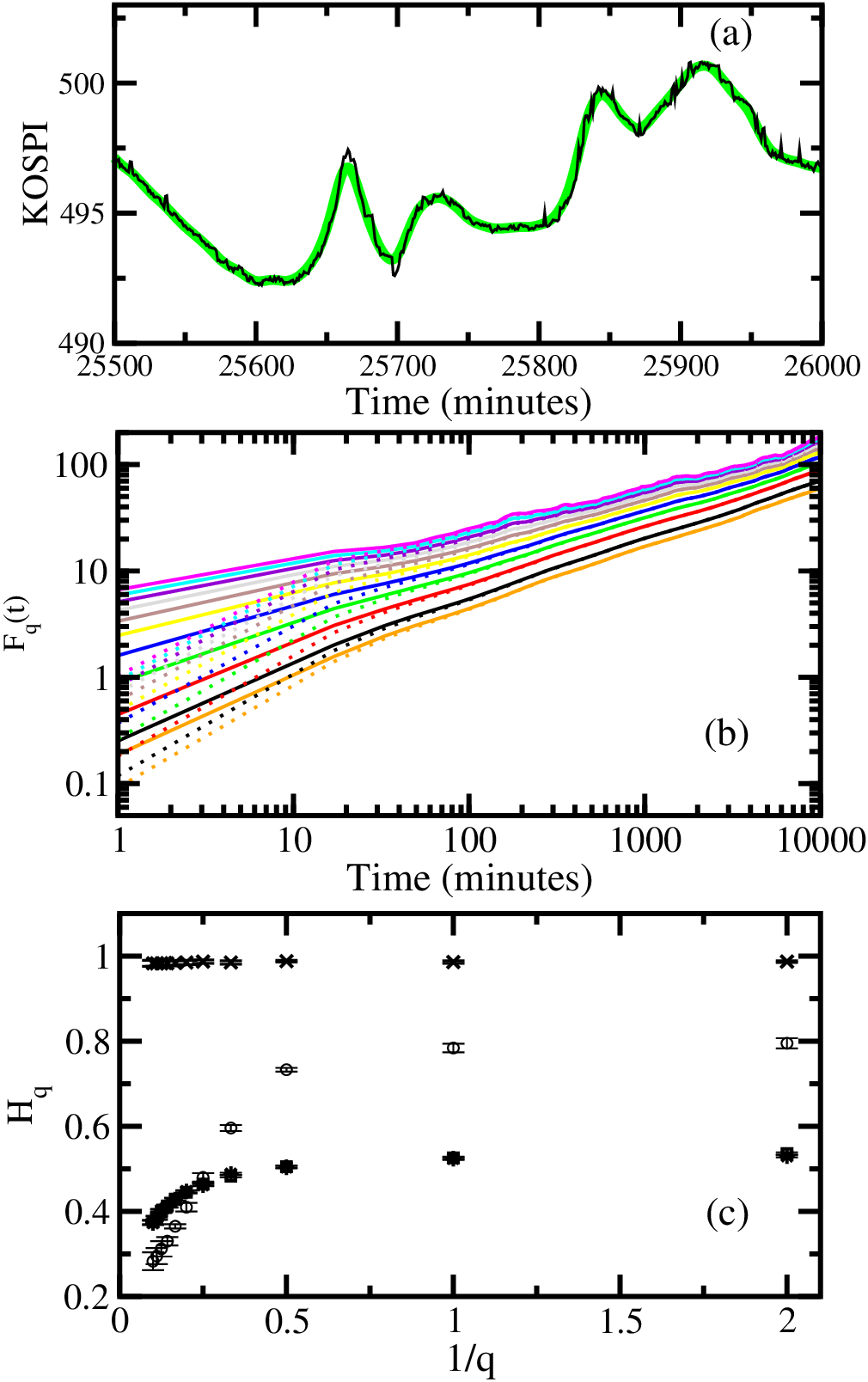}
\caption[0]{ (Color online.)
(a) A portion of the original stock index 
(black)
and the time series convoluted with a Gaussian 
(gray, green online).
(b)
Log-log plots of the generalized height-height correlation function 
$F_q (t)$ versus time
for the original time series (solid curves)
and the time series smoothed by Gaussian convolution (dotted curves)
 for $q=0.5, 1,2, \cdots, 10$
from bottom to top. The same gray scales (colors online) 
correspond to the same values of $q$.
(c) The generalized Hurst exponents $H_q$ versus $1/q$
for the time series smoothed by Gaussian convolution
in the scaling regimes $t<t_c ~(\times)$ and $ t > t_c ~(*)$,
where $t_c \simeq 40$~min.
For easy comparison we also include
the original index in the scaling regimes $t<t_c ~(\circ)$ and $ t > t_c ~(\square)$
(identical to Fig.~1(b)).
}
\label{fig4}
\end{figure}
We also checked the effects of varying the standard deviation in the
Gaussian function between 1~min and 100~min. When the standard deviation is close to
the time interval of the stock-index time series, 
we found that the multifractal behaviors 
are the same as those of the original time series.
When the standard deviation is greater than 4 minutes and less than the
crossover time $t_c = 40$ minutes, the multifractal behavior
is the same as the results obtained with a standard deviation of 5 minutes
in Fig.~4:
$H_q$ is close to one for $t < t_c$, and it depends on $q$ 
for $t>t_c$. 
When the standard deviation is greater than 40 minutes, the crossover 
time increases proportional to the standard deviation. In this case, the
convoluted time series deviate from the global trends of the
original time series. 
Other functions than Gaussian in the convolution, such as a Lorentzian, do
not change the results significantly.
 From these observations, we conclude that the multifractality
in the short-time regime, $t<t_c$, originates from the local fluctuations
of the stock index. The convolution removes the local 
fluctuations and smooths out the variations. However, in the long-time regime,
$t>t_c$, there are no differences  between the convoluted time series
and the original time series in the aspect of multifractality. 
In this regime multifractality may be 
caused by the nonlinearities and the herding dynamics
of the stock index~\cite{EK04,KH02,BT04}. 
The behavior of the GHCF for the convoluted time series removing the
daily jumps is similar to the original convoluted
time series keeping the daily jumps. 
The multifractality in the long-time regime is robust with respect to
smoothing of the time series, and it appears to be an  intrinsic property 
of the stock-market dynamics~\cite{LR04}.

\begin{table}[loc=htbp]
\caption{The generalized Hurst exponents of original and convoluted
stock index for $t<t_c$}
\label{hst}
\begin{tabular}{|c|c|c|}
\hline
q &original stock index $H_q (t<t_c )$ & convoluted stock index $H_q (t < t_c )$ \\ \hline
0.5 & 0.80(1) & 0.987(2) \\ \hline
1 & 0.78(1) & 0.986(3)
 \\ \hline
2 & 0.73(1) & 0.988(2) 
 \\ \hline
3 & 0.60(1) & 0.985(4)
 \\ \hline
4 & 0.48(1) & 0.987(4)
 \\ \hline
5 & 0.41(1) & 0.985(5)
 \\ \hline
6 & 0.37(1) & 0.984(5)
 \\ \hline
7 & 0.33(1) & 0.983(6)
 \\ \hline
8 & 0.31(2) & 0.983(6)
 \\ \hline
9 & 0.29(2) & 0.983(7)
 \\ \hline
10 & 0.28(2) & 0.983(7)
 \\ \hline

\end{tabular}
\end{table}

\section{Conclusion}
In summary, we have studied multifractality of the Korean KOSPI stock index.  
Multifractality is observed in two scaling regimes in the original time
series of the stock index. However, multifractality in the short-time regime
can be removed by convoluting the time series by a Gaussian.
In the long-time regime the multifractality is unchanged
by smoothing the time series.  
We propose that multifractality in the short-time regime is caused by
the local fluctuations in the stock index. However,  multifractality
in the long-time regime appears to be a result of the complex behaviors in the stock market,
such as herding behavior, information outside the market, the
long memory of the volatility, and the intrinsic nonlinear dynamics
of the market. Understanding the origins of  multifractality
in the long-time scaling regime remains an active research topic.
 
\section*{Acknowledgments}
We appreciate useful correspondence with
S. J. Mitchell. 
This work was supported by KOSEF(R05-2003-000-10520-0(2003)) and by U.S. National
Science Foundation Grant Nos. DMR-0120310 and DMR-0240078. Work at Florida State
University was also supported by the School of Computational Science and
the center for Materials Research and Technology.

\newcommand{\jpa}{J. Phys. A}
\newcommand{\jkps}{J. Kor. Phys. Soc.}


\begin{thebibliography}{}

\bibitem{MA99} R. N. Mantegna and H. E. Stanley,  An Introduction to
Econophysics: Correlations and Complexity in Finance, Cambridge University Press,
Cambridge, 1999.
 \bibitem{MA97} B. Mandelbrot, Fractals and Scaling in Finance, Springer,
New York, 1997.
\bibitem{BP00} J. P. Bouchaud and M. Potters, Theory of Financial
Risk, Cambridge University Press, New York, 2000.
\bibitem{SO03} D. Sornette, Phys. Rep.  378 (2003) 1.
 \bibitem{MS95} R. N. Mantegna and H. E. Stanley, Nature  376 (1995) 46.
 \bibitem{BS94} J.-P. Bouchaud and  D. Sornette, J. Phys. I France  4 (1994) 863.
 \bibitem{BA1900} L. Bachelier, Ann. Sci. \'{E}cole Norm. Sup.  3 (1900) 21.
 \bibitem{MA63} B. Mandelbrot, J. Business  36 (1963) 294.
 \bibitem{FA63} E. F. Farma, J. Business  36 (1963) 420.
  \bibitem{GP99} P. Gopikrishnan, V. Plerou, L. A. N. Amaral, M. Meyer, 
and H. E. Stanley, Phys.  Rev. E  60  (1999) 5305.
 \bibitem{GM99} P. Gopikrishan, M. Meyer, L. A. N. Amaral, and H. E. Stanley, 
Eur. Phys. J. B.   3 (1999) 139.
 \bibitem{LK99} Y. Liu, P. Gopikrishnan, P. Cizeau, M. Meyer, C.-K. Peng, 
and H. E. Stanley, Phys. Rev. E  60 (1999) 1390.
 \bibitem{GP00} P. Gopikrishnan, V. Plerou, Y. Liu, L. A. N. Amaral, X. Gabaix,
and  H. E. Stanley, Physica A  287 (2000) 362.
 \bibitem{SA00} H. E. Stanley, L. A. N. Amaral, P. Gopikrishnan, 
and V. Plerou, Physica A  283 (2000) 31.
\bibitem{KK02} H.-J. Kim, I.-M. Kim, and B. Khang, \jkps ~ 40 (2002) 1105.
\bibitem{BC03} G. Bonanno, G. Caldarelli, F. Lillo, and R. N. Mantegna, Phys. Rev. E
 68 (2003) 046130.
\bibitem{NA02} Y. Nakajima, \jkps ~ 40 (2002) 1096.
\bibitem{LE02} S. Y. Park, S. J. Koo, K. E. Lee, J. W. Lee, and B. H. Hong, 
New Physics  44 (2002) 293.
\bibitem{LL04} K. E. Lee and J. W. Lee, \jkps ~ 44 (2004) 668.
\bibitem{GD02} A. Z. G\'{o}rski, S. Drodz, and J. Septh, Physica A  316 (2002) 496.
\bibitem{WH01} B. H. Wang and  P. M. Hui, Eur. Phys. J. B.  20 (2001) 573.
\bibitem{KC04} K. Kim, J.-S. Choi, and S.-M. Yoon, \jkps~ 44 (2004) 643.
\bibitem{SK00} J. A. Skjeltorp, Physica A  283 (2000) 486.
\bibitem{KY03} K. Kim and S. M. Yoon, Fractals  11 (2003) 131.
\bibitem{AR02} J. Alvarez-Ramirez, M. Cisneros, C. Ibarra-Valdez, A. Soriano, and
E. Scalas, Physica A  313 (2002) 651.
\bibitem{XX03} Z. Xin-Tian, H. Xiao-Yuan, and S. Yan-Li, Physica A  333 (2003) 293.
\bibitem{EK04} Z. Eisler and J. Kertesz, Physica A  343 (2004)  603.
\bibitem{BE03} A. Bershadskii, Physica A  317 (2003) 591.
\bibitem{AI02} M. Ausloos and K. Ivanova, Comp. Phys. Comm.  147 (2002) 582.
\bibitem{IA99} K. Ivanova and M. Ausloos, Eur. Phys. J. B  8 (1999) 665.
\bibitem{TP03} A. Turiel and  C. J. Perez-Vicente, Physica A  322 (2003) 629.
\bibitem{KA00} H. Katuragi, Physica A  278 (2000) 275.
\bibitem{BE01} A. Bershadskii, J. Phys. A  34 (2001) L127.
\bibitem{MA03} K. Matia, Y. Ashkenazy, and H. E. Stanley, Europhys. Lett. 
 61 (2003) 422.
\bibitem{XG03} Z. Xu and R. Gencay, Physica A  323 (2003) 578.
\bibitem{BM02} G. M. Buend{\'\i}a, S. J. Mitchell, and P. A. Rikvold, Phys. Rev. E  66 (2002)
046119.
\bibitem{BM03} G. M. Buend{\'\i}a, S. J. Mitchell, and P. A. Rikvold, Microelectronics J. in press.
\bibitem{MI02} S. J. Mitchell, cond-mat/0210239.
\bibitem{MK} P. Meakin, Fractals, scaling and growth far from equilibrium, Cambridge 
University Press, Cambridge, 1998.
\bibitem{NR} W. H. Press, S. A. Teukolsky, W. T. Vetterling, and B. P. Flannery, 
Numerical Recipes in Fortran: The art of scientific computing, Cambridge University Press, Cambridge, 1986, pp 492.
\bibitem{KH02} A. Krawiecki, J. A. Holyst, and D. Helbing, Phys. Rev. Lett.  89 (2002)
 158701.
\bibitem{BT04} M. Barotolozzi and A. W. Thomas, Phys. Rev. E  69 (2004) 046112.
\bibitem{LR04} The performance of our analysis code was checked with a synthetic, monofractal
time series, for which it showed no indication of multifractality.

\end{thebibliography}
\end{document}